\documentclass[twocolumn,english,prl,showpacs,superscriptaddress]{revtex4}
\usepackage{graphicx}
\usepackage{mathrsfs}

\begin{document}
\title{BEC ``level'' for measuring small forces}
\author{S. G. Bhongale}
\email{bhongale@rice.edu}
\affiliation{Department of Physics and Astronomy, MS 61, Rice
University, Houston, TX 77005, USA.}
\author{Eddy Timmermans}
\affiliation{Center for nonlinear Science, LANL, Los Alamos, NM 87545, USA.}

\begin{abstract}
We propose a device that consists of a trapped two-component phase-
separated
Bose-Einstein condensate to measure small forces and
map weak potential energy landscapes. The resolution as well as the
measurement precision of this device can be set dynamically, allowing
measurements at multiple scales.
\end{abstract}

 \date{\today} \maketitle The mesoscopic quantum coherent
matter systems, such as dilute gas Bose-Einstein condensates (BEC's)
and fermion superfluids, created in cold atom traps are unusually
sensitive and clean many-body systems that can be isolated nearly
completely from
the influence of environment, and possess unusual control knobs such
as the ability to vary the inter-particle interaction at will
\cite{feshbach}. This has lead to several suggestions for BEC-probes
that would measure, for example, Casimir forces
\cite{cornell-casimir}, magnetic fields \cite{stamperkurn}, or the
earth's
rotation \cite{stringari}.  In this letter, we discuss the prospect of
creating a mesoscopic sized BEC-object (a BEC ``bubble'') that is free
or nearly free - the effective potential energy experienced by the
bubble's center-of-mass position is nearly constant in space.
Therefore, the bubble's center-of-mass position becomes an
ultra-sensitive measure of any weak external force on the bubble.  At
the same time, if the boundary region of the bubble is `sharp' then
its edge maps out potential energy contours with exquisite accuracy.
The freely floating bubble also provides an intriguing template for
probing mesoscopic quantum behavior.  By crafting shallow
external potentials superimposed upon the flat effective `bubble'
potential, experimentalists could demonstrate many-body
quantum interference, observe quantum Brownian motion and
study many-body quantum tunneling.

\begin{figure}[hb]
  \includegraphics[scale=.55]{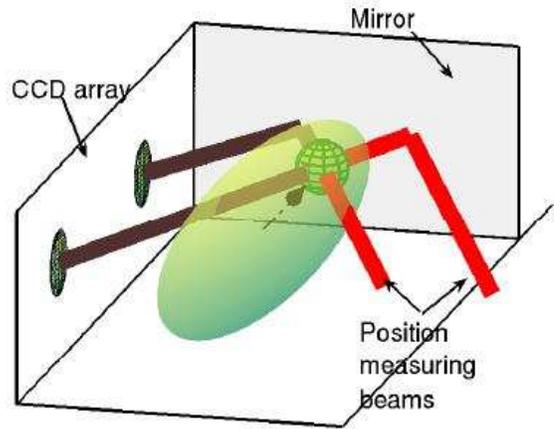}
\caption{Schematic of the proposed BEC ``level'' device. The bubble of
  BEC-B (represented by a sphere) is confined within the BEC-A sea
  (shown by the ellipsoid). The arrow along the axis of the ellipsoid
  is the direction of the external force that causes the bubble to be
  displaced from the center of the ellipsoid. This displacement can be
  measured by a laser configuration as shown in the figure.}
\label{cartoon}
\end{figure}

The actual bubble we propose to use is a trapped phase-separated BEC
of `B' atoms floating within a simultaneously trapped, larger BEC of
`A' atoms \cite{timmermans,fermiseparation} as shown in the schematic
of Fig.~\ref{cartoon}.  The extent of the phase separated BEC-B bubble
is confined by the BEC-A sea in which it floats.  By carefully
choosing the trapping potentials and interaction strengths, the
trapping, buoyancy, and surface tension forces on the bubble can
cancel, resulting in a freely floating mesoscopic BEC object near the
middle of the trap.  An additional increase of the inter-species
interaction strength decreases the size of the interface boundary
region between the BEC's, sharpening the `edge' of the bubble.  Now
the bubble's position is extraordinarily sensitive to the magnitude of
a small perturbing external force.  Thus, the position of the bubble
becomes a measure of the magnitude of this force. Moreover,
experimental methods based on imaging the shadow of the bubble as
depicted in Fig.~\ref{cartoon}, can accurately measure the bubble
displacement.  Such techniques have proved to be yield exquisitely
precise position measurements elsewhere \cite{imaging}.

Conceptually, the proposed weak force detector resembles a ``level'' -
a device that determines the horizontal direction by detecting the
cancellation of the gravitational force component in the direction
in which a bubble in a fluid-filled tube can move.  At the point of
cancellation, the bubble remains stationary in the middle of the tube.
If the effective potential of the phase separated BEC bubble is exactly
flat, then the actual ground state of the system is a superposition
of states in which the bubble is located in different positions.
If the bubble behaves as a macroscopic object, then once its position
is measured, it remains localized, `pinned' by interactions
with the environment - an essential ingredient of the breaking of
translational symmetry.  If the environment interactions are
sufficiently weak
and in the presence of a shallow double well potential, the
bubble can behave as a quantum object, tunneling between the wells
(the bubble's surface tension may prevent single particle
tunneling) or interfering with itself if it can follow
different paths to reach the same final position. Such experiments
that reveal the quantum properties of the `bubble' can be carried out
on a `chip' \cite{chip} (near a surface). The main source of decoherence
would be the excitation of phonon modes in the surrounding BEC-sea.

For the discussion in this letter, we won't require the extreme
level of control necessary for the observation of mesoscopic quantum
behavior. Instead, we assume that external interactions are present
(though small) and that the BEC `bubble' acts as a classical object.
Even in that case, the bubbles can probe genuine quantum
effects, for example, the position of two bubbles each trapped in one
well of a double well potential would be modified from their
equilibrium position if they experience a mutual attraction, induced
by Casimir-like force arising from quantum fluctuations of the
surrounding BEC \cite{roberts}.

We illustrate the cancellation of bubble forces in the case of a
'small bubble of vanishing surface tension' - the simplest limit.
In that case, the size of the bubble is sufficiently large to neglect
the
surface energy, but remains small compared to the length scale on
which the trapping potentials $V_{A}({\bf r})$ and $V_{B}({\bf r})$
experienced by the `A' and `B' atoms respectively, vary.  Hence, we
can replace $V_{B}({\bf r}) \approx V_{B}({\bf R})$ where $ {\bf R}$
denotes the bubble's center-of-mass position.  We also assume that the
BEC gas is dilute and that we can describe the inter-particle
interactions by the customary contact potentials: $v_{A(B)}({\bf
  r}-{\bf r}') = \lambda_{A(B)} \delta({\bf r}-{\bf r}')$ for the
mutual interactions of like bosons `A' (or `B') and $v_{AB}({\bf
  r}-{\bf x}) = \lambda \delta({\bf r}-{\bf x})$ for the mutual
interactions of unlike bosons.  In a three-dimensional BEC gas, the
contact interactions lead to local pressures within the single
condensate regions of average density $\rho_{A(B)}({\bf r})$ equal to
$[\lambda_{A(B)}/2] \rho_{A(B)}^{2} ({\bf r})$.  If we neglect the
surface tension of the bubble, then the condition of equilibrium
requires the inside and outside bubble pressures to cancel, i.e. $P=
\lambda_{A} [\overline{\rho}_{A}({\bf R})]^2/2 =
\lambda_{B}[\rho_{B}({\bf R})]^2/2 $, where $\overline{\rho}_{A}$
denotes the density of an all `A' BEC of the same chemical potential
as the actual `A' BEC which surrounds the B-bubble. Hence, the inside
`bubble' density is equal to $\rho_{B}({\bf R}) =
\sqrt{\lambda_{A}/\lambda_{B}} \overline{\rho}_ {A}({\bf R})$.  We
write the ground state energy of the trapped BEC `A' which has the `B'
bubble of $N_{B}$ particles immersed as the energy $\overline{E}_{A} $
of the all `A' BEC that has the same chemical potential as the one
with the bubble, and an integral over the `B' bubble volume
$\Omega_{B}$ centered on ${\bf R}$.  Consistent with the assumption of
slowly varying trapping potentials and the neglect of the bubble
surface energy, we approximate the energy densities as $V_{A(B)}({\bf
  r}) \rho_{A(B)}({\bf r}) + [\lambda_{A(B)}/2] \rho_{A(B)}^{2}({\bf
  r})$ and we obtain
\begin{eqnarray}
E&=& \overline{E}_{A} + \int_{\Omega_{B}} \left[ V_{B}({\bf r})
  \rho_{B}({\bf r}) + \lambda_{B} \rho_{B}^{2}({\bf r})/2
  \right] \nonumber\\
&& \; \; \; \; \; - \left[ V_{A}({\bf r})
  \overline{\rho}_{A}({\bf r}) + \lambda_ {A}
  \overline{\rho}_{A}^{2}({\bf r})/2 \right] d{\bf r}\nonumber\\
&\approx&
\overline{E}_{A} + N_{B} \left[ V_{B}({\bf R}) -
  \sqrt{\lambda_{A}/\lambda_{B}} V_{B} ({\bf R}) \right] \; \;
\nonumber ,
\end{eqnarray}
consistent with an effective potential energy per B-particle equal to
$V_{\text{eff},B}({\bf R}) = V_{B}({\bf R}) -
\sqrt{\lambda_{B}/\lambda_{A}} V_{A}({\bf R})$.  Now, if the trapping
potentials for A and B particles are simply proportional, $V_{A} ({\bf
  r}) = \alpha V_{B}({\bf r})$, and we carefully tune the
proportionality constant $\alpha$ to $\alpha \rightarrow
\sqrt{\lambda_{A}/\lambda_{B}}$ then the effective potential as seen
by the bubble vanishes and it floats freely inside the BEC-A.

This value, $\alpha = \sqrt{\lambda_{A}/\lambda_{B}}$ is also the
proportionality constant at which the bubble's position becomes
highly sensitive to any small additional external force.  We
illustrate this for the case of a harmonic trapping potentials,
$V_{B,\text{trap}} ({\bf r}) =(1/2)\left[ K_{x} x^{2} + K_{y} y^{2} +
  K_{z}z^{2} \right]$, $V_{A,\text{trap}}({\bf r}) =\alpha
V_{B,\text{trap}}({\bf r})$, to which a differential external force
${\bf F}=\gamma \hat{z}$ (i.e. difference of force experienced by the
A- and B- atoms) adds an approximately linear term to the B-potential
$V_{B}({\bf r}) = V_{B,\text{trap}}({\bf r})- \gamma z$, so that
\begin{eqnarray}
V_{\text{eff},B}({\bf r}) &=& \left( 1 - \alpha
\sqrt{\lambda_{B}/\lambda_{A}} \right)V_{B,\text{trap}} - \gamma z
\\ &=& [(1 - \alpha
  \sqrt{\lambda_{B}/\lambda_{A}})/2]\times\nonumber\\ &&\left[ K_{x}
  x^{2} + K_{y} y^{2} + K_{z} (z-\Delta)^{2} - K_{z} \Delta^ {2}
  \right] \; \; ,
\end{eqnarray}
centered around a position that is displaced along the $z$-axis from
the former trap middle by a distance $\Delta=\gamma[K_z \left( 1 -
  \alpha \sqrt{\lambda_{B}/\lambda_{A}} \right)]^{-1}$.  Hence, in
equilibrium, the measurement $\Delta$ provides a direct measurement of
the magnitude of the force, $|{\bf F}|=\gamma$.

The accuracy with which the bubble's position can be measured depends
on the sharpness of its edge. In the edge region if we assume the
respective BEC densities to vary as $\rho_B(\epsilon)=\rho_B({\bf
  R})(\delta-\epsilon)/\delta$ and $\rho_A(\epsilon)=\rho_A({\bf
  R})\epsilon/\delta$, where $0\le\epsilon\le\delta$ is normal to the
interface area ${\mathcal A}$, then the resulting kinetic energy
contribution can be approximated by $E_{\text{kin},\delta}=4 {\mathcal
  A}P[l_A^2+l_B^2]/\delta$, where
$l_{A(B)}=\hbar/\sqrt{4m_{A(B)}n_{A(B)}\lambda_{A(B)}}$ is the
coherence length of BEC-A(B). We also find the surface interaction
energy to be approximately equal to $E_{\text{int},\delta}={\mathcal
  A}P(\lambda-\sqrt{\lambda_A\lambda_B})\delta/3$. Thus the
equilibrium `edge' width obtained by minimizing this additional
surface energy
$E_{\text{edge}}=E_{\text{kin},\delta}+E_{\text{int},\delta}$ is given
by $\overline{\delta}=
\sqrt{12(l_A^2+l_B^2)/(\lambda/\sqrt{\lambda_A\lambda_B}-1})$
\cite{timmermans}.  Therefore the `edge' sharpness can be increased by
increasing the interaction strength between unlike bosons by means of
an interspecies scattering resonance
\cite{interspeciesfeshbach}. This increases the importance of the
surface
energy, requiring either a description that includes surface tension
explicitly, or a numerical solution of the coupled Gross-Pitaevskii (GP)
like equations.  Given the lack of space, we choose the latter option
and solve the numerical equations by evolving the GP-dynamics in
imaginary time. Motivated by recent experiments with elongated BEC's
\cite{schmiedmayer} created by focusing a singe laser beam, we consider
a quasi-1D configuration with $\omega_z$ and $\omega_{\perp}$, the
trapping frequency in the axial and transverse direction respectively,
and $\xi=\omega_z/\omega_{\perp}<<1$ for both species of bosons.

\begin{figure}[t]
  \includegraphics[scale=.48]{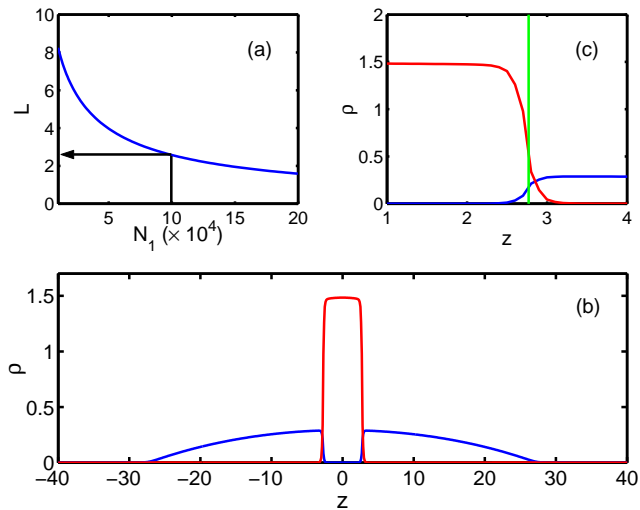}
  \caption{Figure comparing an analytical
    calculation that neglects the surface tension and the numerical
GP solution. In (a) the
    half axial length, $L$, of the bubble is plotted as a function
of $N_1$ for
    $N_2=2\times10^4$, $\alpha=0.9$, $\xi=0.01$, $\beta=0.9$, and
    $a_1=0.001$, (b) the density profile obtained numerically for
    $N_1=10^5$ is shown, and (c) shows the magnified view of the
    interface region in (b). The green line indicates the boundary
    between BEC-A and -B obtained in (a).  }
\label{combined}
\end{figure}

The wavefunctions, $\phi_A$ and $\phi_B$ of the two interacting
condensates evolve according to the coupled Gross-Pitaevskii (GP)
equations
\begin{eqnarray}
i\dot{\phi}_{A(B)}= \Big[\frac{-\nabla^2}{2}- \mu_{A(B)}+V_{A(B)}({\bf
    r})\hspace{2.5cm}&&\nonumber\\
+\lambda_{A(B)}N_{A(B)}|\phi_{A(B)}|^2\Big]\phi_{A(B)}
+\lambda N_{B(A)}|\phi_{B(A)}|^2\phi_{A(B)}&&\nonumber
\end{eqnarray}
where the mass of `A' and `B' atoms is assumed to be same, $V_A({\bf
  r})=\alpha V_B({\bf r})=\alpha(x^2/2\xi^2+y^2/2\xi^2+z^2/2)$, all
the physical quantities are scaled in axial harmonic oscillator units,
and the condensate wavefunctions $\phi_{A(B)}$ are normalized to
unity. We work in the phase separated regime by using
$\lambda/\sqrt{\lambda_A\lambda_B}=1.1$. The density profile obtained
numerically is plotted in Fig.~\ref{combined}. In the same figure we
also plot the value of the half axial length, $L$, obtained by
neglecting surface tension and taking the Thomas-Fermi
approximation. For treating the quasi-1D situation considered
here, we use a modified local density ansatz given by
$\mu_{A(B)}(z)=\mu_{A(B)}-V_{A(B)}(z)$ and
$\mu_{A(B)}(z)=\sqrt{\alpha/\xi^2}\sqrt{1+4a_{A(B)}\rho_{A(B)}(z)}$
where $a_{A(B)}$ is the two-body $s$-wave scattering length for atoms
`A'(`B') \cite{gerbier} which determines the interaction strength by the
relation $\lambda_{A(B)}=4\pi\hbar^2a_{A(B)}/m$. Figure
\ref{combined}.(c) shows excellent agreement between $L$ and the
numerically obtained location of the bubble edge.

\begin{figure}[t]
  \includegraphics[scale=.45]{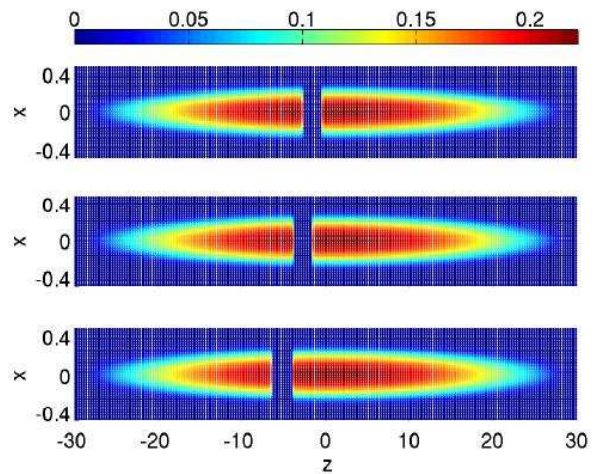}
\caption{Density of BEC-A is shown as a surface plot for
  $\gamma=0.1$. The subplots top, middle, and bottom correspond to
  $\alpha\sqrt{\lambda_B/\lambda_A}=0.90, 0.93$ and $0.95$
  respectively. The BEC-B bubble is located within BEC-A where
  $\rho_A({\bf r})=0$.}
\label{levelposition}
\end{figure}

We demonstrate the working of the BEC ``level'' by considering an
external force in the axial direction ${\bf F}=\gamma \hat{z}$. As
argued before, we tune the inter-atomic interactions such that
$\alpha\sqrt{\lambda_B/\lambda_A}\rightarrow 1^-$ and
$\lambda>\sqrt{\lambda_A\lambda_B}$. The corresponding ground state
density profile obtained numerically is shown in
Fig.~\ref{levelposition}. Here we see that the BEC-B bubble is
displaced from the center of the trap by a distance
$\Delta\approx\gamma/(1-\alpha\sqrt{\lambda_B/\lambda_A})$.  The
boson-boson interaction strength as well as the trapping frequencies
can be precisely tuned in current cold-atom experimental setups
resulting in a large displacement of the bubble and thus allowing for
the precise measurement of $|{\bf F}|=\gamma$.

\begin{figure}[t]
  \includegraphics[scale=.49]{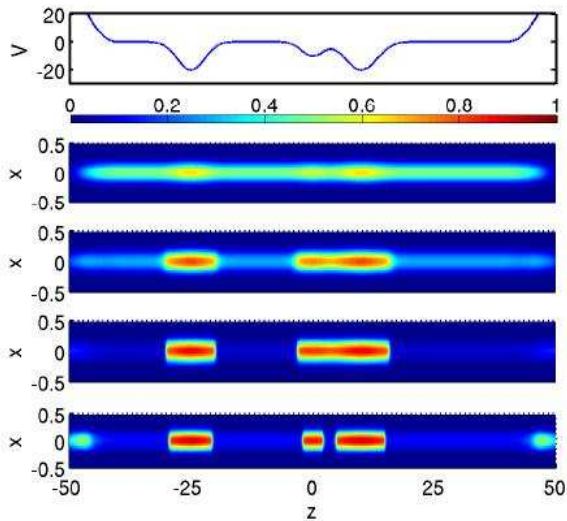}
\caption{Density profile of the quasi-1D BEC-B for different values of
  $N_A$. The topmost figure shows the trapping potential while the
  remaining four plots (from top to bottom) correspond to $N_A=0,
  1000, 1500, 2000$.}
\label{diffn}
\end{figure}

While the above discussion was related to determining the force
(gradient of the external potential), the proposed BEC ``level'' can
also be used for mapping accurate potential energy contours. In fact
such a measurement can be achieved using a single BEC as
demonstrated in \cite{schmiedmayer}. This experiment related the
observed BEC-density profile to the variation of the external
potential. The experimental accuracy of the potential was a few Hz
with a spatial resolution of few
microns. One limitation of the potential measurement is set by the
magnitude of the chemical potential. One cannot arbitrarily decrease
the chemical potential or the density of the BEC will drop below what
is necessary for imaging the atoms or, if the inter-particle
interaction is
lowered, the coherence length of the BEC will increase to limit the
spatial resolution. In contrast, the phase separated BEC configuration
discussed in the present letter not only ensures large density inside
the bubble allowing good imaging, but an edge width that can be
reduced below typical BEC-coherence lengths allowing increased spatial
resolution. We demonstrate this principle by considering an axially
varying
potential $V_{\text{trap}}(z)$ as shown in the topmost subplot of
Fig.~\ref{diffn}. In the same figure, the remaining subplots, show the
density profile of the BEC-B as a function of the total number of `A'
atoms, $N_A$. Here we see that as $N_A$ is increased, the equilibrium
size of the BEC-B bubble decreases. This continues till a point is
reached when the BEC-B bubble is small enough and is localized in the
valleys of the potential landscape. Moreover, as mentioned earlier,
the edge sharpness can be improved by modifying the inter-species
interaction strength $\lambda$.  Thus we see that the proposed BEC
``level" device can map potential energy contours arising from,
for example, surface magnetic fields with exquisite precision.

In conclusion, we proposed a device for measuring small forces
(gradient of potential) by using a phase separated two component BEC
configuration. The working of this device is analogous to that of a
level, so that we named it a ``BEC-level''. This device can
also map out potential energy contours with exquisite accuracy.  This
device could, for instance, map potential energy variations experienced
by atoms near the surface of a metal caused by fluctuations of
the surface fields. The BEC-level provides several knobs that can
be tuned, allowing for measurements to be carried out at multiple
levels of
accuracy.

While we have only exploited the bubble as a classical object, its
possible range of explorations includes mesoscopic quantum behaviour
such as
many-body tunneling, quantum interference, and quantum Brownian
motion. Along with the device proposed in this letter, the BEC-level
can be used as a template for probing quantum many-body behaviour.

S.B acknowledges financial support from the W. M. Keck Program in
Quantum Materials at Rice University. E.T. acknowledges support from
the laboratory directed research and development (LDRD) program.


\begin{thebibliography}{99}
\bibitem{feshbach}C. Chin et al., Science, {\bf 305}, 1128 (2004);
  C. A. Regal, M. Greiner, and D. S. Jin, Phys. Rev. Lett. {\bf 92},
  040403 (2004); M. W. Zwierlein et al., cond-mat/0403049 (2004);
  T. Bourdel et al., Phys. Rev. Lett. {\bf 93}, 050401 (2004); K.M. O'
  Hara et al., Science {\bf 298}, 2179 (2002); R. Hulet, presentation
  in the Quantum Gas Conference at the Kavli Institute for Theoretical
  Physics, Santa Barbara, CA, May 10-14 (2004).
\bibitem{cornell-casimir}J. M. Obrecht, R. J. Wild, M. Antezza,
  L. P. Pitaevskii, S. Stringari, and E. A. Cornell,
  Phys. Rev. Lett. {\bf 98}, 063201 (2007).
\bibitem{stamperkurn} M. Vengalattore, J. M. Higbie, S. R. Leslie,
  J. Guzman, L. E. Sadler, and D. M. Stamper-Kurn,
  Phys. Rev. Lett. {\bf 98}, 200801 (2007).
\bibitem{stringari}S. Stringari, Phys. Rev. Lett. {\bf 86}, 4725
  (2001).
\bibitem{timmermans}Eddy Timmermans, cond-mat/9709301 ; Eddy
  Timmermans, Phys. Rev. Lett. {\bf 81}, 5718 (1998); J. Stenger,
  S. Inouye, D. M. Stamper-Kurn, H. -J. Miesner, A. P. Chikkatur, and
  W. Ketterle, Nature {\bf 396}, 345 (1998); D. S. Hall,
  M. R. Matthews, J. R. Ensher, C. E. Wieman, and E. A. Cornell,
  Phys. Rev. Lett. {\bf 81}, 1539 (1998) .
\bibitem{fermiseparation} Such phase separation occurs in trapped
  Fermi-gas systems as well: G. B. Partridge, W. Li, R. I. Kamar,
  Yean-an Liao, and R. G. Hulet, Science {\bf 311}, 503 (2006);
  T. N. De Silva and E. J. Mueller, Phys. Rev. A {\bf 73} 051602
  (2006); K. B. Gubbels, M. W. J. Romans, and H. T. C. Stoof,
  Phys. Rev. Lett. {\bf 97}, 210402 (2006).
\bibitem{imaging} S. Schneider, A. Kasper, Ch. vom Hagen,
  M. Bartenstein, B. Engeser, T. Schumm, I. Bar-Joseph, R. Folman,
  L. Feenstra, and J. Schmiedmayer, Phys. Rev. A {\bf 67} 023612
  (2003), D. M. Harber, J. M. Obrecht, J. M. McGuirk, and
  E. A. Cornell, Phys. Rev. A {\bf 72}, 033610 (2005).
\bibitem{chip} S. Aubin, S. Myrskog, M. H. T. Extavour, L. J. Leblanc,
  D. Mckay, A. Stummer, and J. H. Thywissen, Nature {\bf 2}, 384
  (2006); Ying-Ju Wang, D. Z. Anderson, V. M. Bright, E. A. Cornell,
  Q. Diot, T. Kishimoto, M. Prentiss, M. Saravanan, S. R. Segal, and
  S. Wu, Phys. Rev. Lett. {\bf 94}, 090405 (2005); Y. Shin, C. Sanner,
  G. -B. Jo, T. A. Pasquini, M. Saba, W. Ketterle, D. E. Pritchard,
  M. Vengalattore, and M. Prentiss, Phys. Rev. A {\bf 72} 021604
  (2005).
\bibitem{roberts} D. C. Roberts and Y. Pomeau, Phys. Rev. Lett. {\bf 95}, 145303 (2005).
\bibitem{interspeciesfeshbach}M. Zaccanti, C. D'Errico,
  F. Ferlaino, G. Roati, M. Inguscio, and G. Modugno, Phys. Rev. A
  {\bf 74}, 041605 (2006); and references therein.
\bibitem{gerbier}F. Gerbier, Europhys. Lett. {\bf 66}, 771 (2004).
\bibitem{schmiedmayer}P. Kr\"{u}ger, S. Wildermuth,
  S. Hofferberth,M. Andersson, S. Groth, I. Bar-Joseph, and
  J. Schmiedmayer, J. Phys. {\bf 19}, 56 (2005).




\end{thebibliography}
\end{document}